\documentclass[twocolumn,amsmath,amssymb]{revtex4-1}

\usepackage{hyperref}
\usepackage{graphicx}
\usepackage{bm}
\usepackage{xcolor}

\begin{document}

\title{Generalizing Thiele Equation}

\author{Bom Soo Kim}
\affiliation{%
	Department of Mathematics and Physics, University of Wisconsin-Parkside,  Kenosha WI 53141,
	USA
}%

\date{\today}

\vspace{0.5in}
\begin{abstract}
	We generalize the Thiele equation with a transverse velocity to the skyrmion motion described by the collective coordinate of magnetization vector. It is applied to investigate significant disparity in the existing data sets of skyrmion and antiskyrmion Hall angles. Our analysis further reveals interesting differences of these Hall angles near the angular momentum compensation point. We identify a possible physical quantity that is responsible for the disparity. 
\end{abstract}

\maketitle

\noindent 
Thiele \cite{Thiele} described the steady state motion of domain wall or skyrmion by parameterizing the unit magnetization vector $\vec n$ as 
 \begin{align} 
	n_i = M_i ( x_j - X_j)/M_s \;, 
\end{align}
where $M_s = |\vec M| $ is a saturation magnetization, and $x_i$ the field position. Roman indices $i, j$ represent the spatial vector components with repeated ones summed over. The collective coordinate $X_j$ is parametrized as 
 \begin{align} 
	X_i = v_i t 
\end{align}
to describe the center of domain wall or skyrmion, which moves with a velocity $v_i$ as time $t$ ticks. 

The resulting Thiele equation has the form 
 \begin{align} \label{ThieleEQOriginal}
	\mathcal G_{ij} v_j + \alpha \mathcal D_{ij} v_j + F_i = 0 \;,
\end{align}
where  $\alpha$ is a damping parameter, and 
 \begin{align} \label{GFactor} 
	\begin{split}
	\mathcal G_{ij} &= \frac{M_s}{\gamma_0} \epsilon_{lmn}  \large\int d^2 x ~n_l (\partial_i n_m) (\partial_j n_n)   \;, \\
	\mathcal D_{ij} &= -\frac{M_s}{\gamma_0}  \large\int d^2 x ~(\partial_i n_k) (\partial_j n_k) \;, \\
	F_i &= -M_s \int d^2 x (\partial_i n_j) H_{\text{eff},j} \;.
	\end{split}
\end{align}
Here $\gamma_0$ is gyromagnetic ratio, $\mathcal G_{ij} v_j$ is the Magnus force, $\mathcal D_{ij} v_j$ is total dissipative drag force, and $ F_i $ is total external force that includes forces due to effective magnetic field and various spin torques. Internal forces due to anisotropy and exchange energies, internal demagnetizing fields, magnetostriction do not contribute \cite{Thiele}. 

Since its publication on 1973, the Thiele equation has been extensively used to describe the steady state motion of magnetic structures such as domain walls or skyrmions. Note that the Magnus force is proportional to the skyrmion charge $Q$, $\mathcal G_{ij} \propto \epsilon_{ij} Q = \epsilon_{ij}  \large\int d^2 x ~  \epsilon_{lmn} n_l (\partial_x n_m) (\partial_y n_n)$ in a 2 dimensional film geometry with coordinates $(x,y)$.

Here we propose to generalize the Thiele equation \eqref{ThieleEQOriginal} with an additional transverse velocity in addition to $v_i$ of the collective coordinate as \cite{Kim:2020piv}
 \begin{align} \label{SkyrmionCM}
	X_i = v_i t + R \epsilon_{ij} v_j t \;,
\end{align}
where the parameter $R$ represents the strength of the transverse velocity compared to the original one. This generalization has been introduced in \cite{Kim:2020piv}. Here we further examine its consequences near the angular momentum compensation point in the context of ferrimagnets. 

The Thiele equation can be derived from the Landau-Lifshitz-Gilbert (LLG) equation \cite{LandauFluidMechanics}\cite{GilbertEq}
\begin{align}
	\partial_t \vec M = - \gamma_0 \vec M \times \vec H_{\text{eff}} + \frac{\alpha}{M_s} \vec M  \times \partial_t \vec M \;. 
\end{align}
As $M_i$ and $\partial_t M_i$ are orthogonal to each other (due to the normalized magnetization vector, $ \vec n^2 =1$), one can show the equation 
 \begin{align} \label{LLGEQ}
	- \frac{\epsilon_{jkl} M_k \partial_t M_l}{\gamma_0 M_s^2}  - \alpha \frac{ \partial_t M_j }{\gamma_0 M_s} + \tilde \beta M_j + H_{\text{eff},j} = 0 
\end{align}
is equivalent to the LLG equation. Explicitly, it can be checked by multiplying $ -\epsilon_{jik} M_k$ to \eqref{LLGEQ}, summing over $j$, and renaming the indices. The third term in \eqref{LLGEQ} does not contribute below as its coefficient is fixed as $\tilde \beta = - M_j H_j/M_s^2$ that can be verified by multiplying $M_j$ to \eqref{LLGEQ}. 

Thiele equation has been used recently, for example, to compute the velocity of various topological spin structures such as  ferromagnetic skyrmion-based logic gates and diodes \cite{ThieleEQAdded1}, antiferromagnetic skyrmion-based oscillators \cite{ThieleEQAdded2} and antiferromagnetic bimerons \cite{ThieleEQAdded3}.

To derive the Thiele equation with the generalization \eqref{SkyrmionCM}, we multiply $-\partial M_j/ \partial x_i $ to \eqref{LLGEQ}. The third term drops out as $ M_j \partial_i M_j = 0$. Note the time derivative $\partial_t \vec M$ has an additional contribution due to the second term in \eqref{SkyrmionCM}.  
\begin{align}
	\partial_t M_i = (v_j + R \epsilon_{jk} v_k) \partial_j M_i \;. 
\end{align}
By integrating over a (skyrmion) volume, one arrives at  
 \begin{align} \label{ThieleEQ}
	\mathcal G_{ij} (v_j + R \epsilon_{jk} v_k) + \alpha \mathcal D_{ij} (v_j + R \epsilon_{jk} v_k) + F_i = 0 \;.
\end{align}
Two coefficients $\mathcal G_{ij}, \mathcal D_{ij}$ and the force term $F_i$ are given in \eqref{GFactor}. This is the Thiele equation generalized with the transverse velocity. Hereafter, we use the same notations of \eqref{ThieleEQ} after multiplying $\gamma_0/M_s $, for example, $(\gamma_0/M_s) F_i \to F_i  $.

It is interesting to see some general features of the new contributions. First, $\mathcal G_{ij} = G \epsilon_{ij}$, antisymmetric with the two indices $ij$. Thus 
\begin{align}
	\mathcal G_{ij} (v_j + R \epsilon_{jk} v_k) = G \epsilon_{ij} v_j - G R v_i \;.
\end{align}
Thus the new term $	\mathcal G_{ij}  R \epsilon_{jk} v_k$ actually increase or decrease the longitudinal velocity $v_i$ depending on the sign of $GR$, while the original term $\mathcal G_{ij} v_j$ is transverse to $v_j$ that is the Magnus force. 

Second, we decompose the drag tensor into symmetric and antisymmetric parts as $ \mathcal D_{ij} = D_s \delta_{ij} + D_a \epsilon_{ij} $. The symmetric part gives 
 \begin{align}  \label{DDecomposition}
	\alpha D_s v_i  + \alpha D_s R \epsilon_{ik} v_k \;.
\end{align}
The new contribution $ \alpha D_s R \epsilon_{ik} v_k $ modifies the transverse motion added to the contribution $ G \epsilon_{ij} v_j$, while $\alpha D_s v_i $ is the contribution Thiele derived. 

On the other hand, the antisymmetric part is given by  
 \begin{align} 
	\alpha D_a \epsilon_{ij} v_j  - \alpha D_a R v_i \;.
\end{align}
This has the same structures of the first term in \eqref{ThieleEQ}. Thus these can be combined as $ G \to G + \alpha D_a$. 
Then our result \eqref{ThieleEQ} can be recast as 
 \begin{align} \label{ThieleEQComponents}
	\begin{split}
	&\big[\alpha D_s - (G+\alpha D_a)R \big] v_i  \\
	&+ \big[ (G+\alpha D_a) + \alpha D_s R \big] \epsilon_{ij} v_j + F_i = 0 \;.
	\end{split}
\end{align}
Thus we see that both the original longitudinal and transverse velocity components are modified by $R$.  

Using this form \eqref{ThieleEQComponents} in 2 dimensions $(x,y)$, it is straightforward to compute the Hall angle. Without loss of generality, we set the force along the $x$ coordinate. Then the equation along $y$ direction gives $[\alpha D_s - (G+\alpha D_a)R ] v_y  = [ (G+\alpha D_a) + \alpha D_s R ]  v_x $. Then the Hall angle is given by  
 \begin{align} \label{HallAngleGeneral}
	\begin{split}
		&\tan \theta_H = \frac{v_y}{v_x} = \frac{ (G+\alpha D_a) + \alpha D_s R}{\alpha D_s - (G+\alpha D_a)R } \;.
	\end{split}
\end{align}
This is our generalization of the Hall angle in the presence of the new transverse effect \eqref{SkyrmionCM}. It does not depend on details of the external force. Note that this Hall angle depends on the parameter $R$ in two different ways, which are related to the transverse contribution $ \alpha D_s R \epsilon_{ik} v_k $ and the longitudinal ones $ - (G +  \alpha D_a )R v_i$. A simpler version of \eqref{HallAngleGeneral} appeared previously in \cite{KimBook}\cite{Kim:2020piv}. 

Without the new transverse effect ($R=0$), the Hall angle \eqref{HallAngleGeneral} reduces to 
 \begin{align} \label{HallAngleOriginal}
	\begin{split}
		&\tan \theta_H = \frac{v_y}{v_x} = \frac{ G+\alpha D_a}{\alpha D_s  } \;.
	\end{split}
\end{align}
We further checked $ D_a = 0$ for the stable positive- and negative-charge skyrmions with $Q= \pm 4\pi $ \cite{Jiang2017}. Thus their Hall angles are the same with opposite signs. 

We note that our convention for skyrmion topological charge has an additional $4\pi$ \cite{footnoteSkyrmionCharge} compared to the definition given in \cite{SKCharge}. There are  related confusions in literature \cite{Han} as mentioned in \cite{footnoteSkyrmionCharge}. 

After describing continuous skyrmion model, we look into basic aspects of the generalized Thiele equation \eqref{ThieleEQ} and investigate the positive- and negative-charge skyrmion Hall angles using \eqref{HallAngleGeneral}.  

\begin{figure}[t!]
	\begin{center}
		\includegraphics[width=0.45\textwidth]{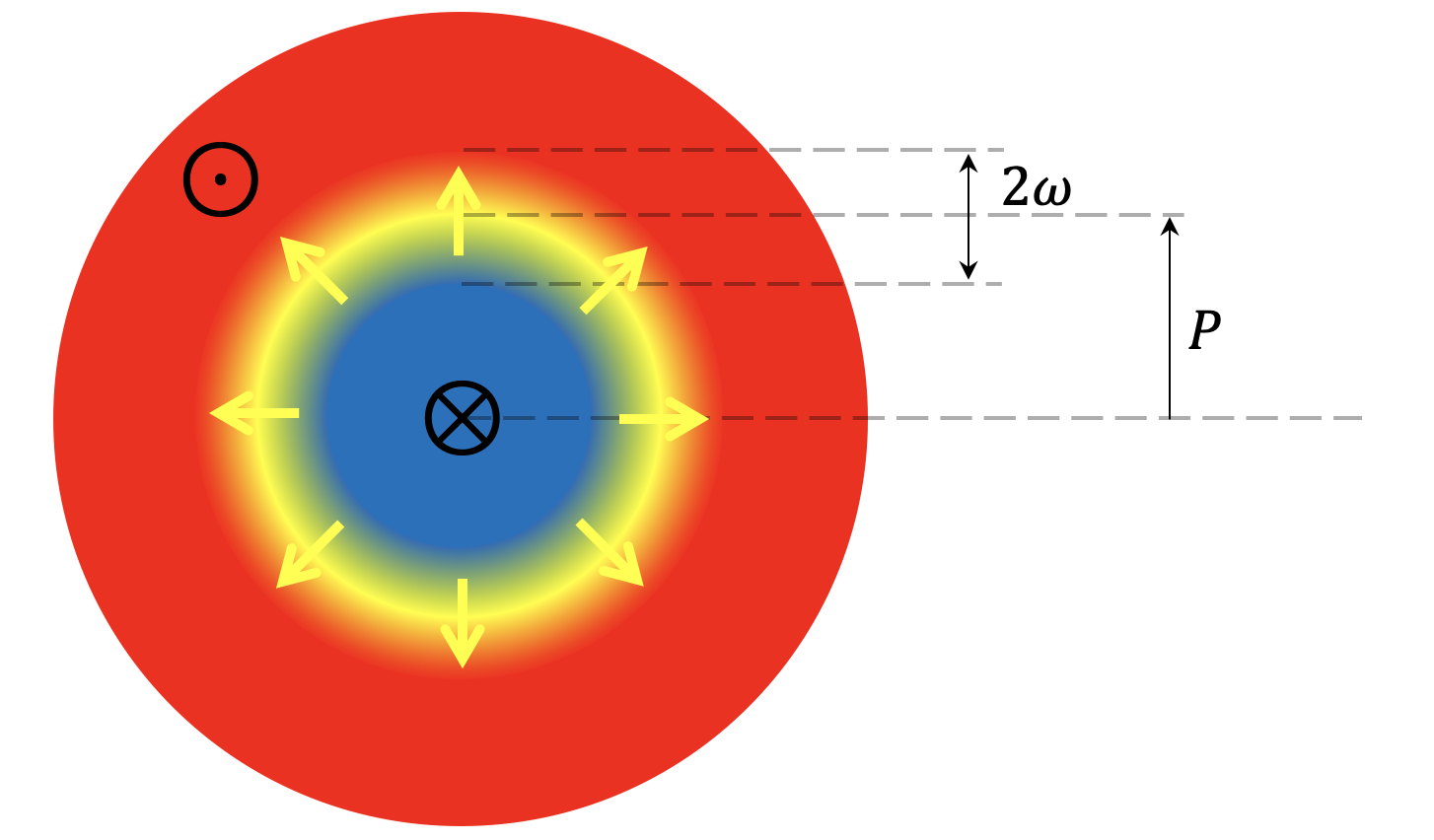} 
		\caption{\footnotesize\small  Skyrmion model with spin‐up (red) in the outer region, spin‐down (blue) in the inner region, and domain wall (yellow, located at $\rho=P$ with width $2\omega$) interpolating between them.  \vspace{-0.2in} 
		}
		\label{fig:SkyrmionModel}
	\end{center}
\end{figure}

\noindent {\bf Continuous skyrmion model}  \\  
The constraint $\vec n^2 =1$ allows us to parametrize the vector $\vec n = \sin \Theta  \cos \Phi  \hat \rho + \sin \Theta  \sin \Phi  \hat \phi + \cos \Theta  \hat z$ with two dimensional coordinates $(\rho= (x^2 + y^2)^{-1/2}, \phi = \tan^{-1} (y/x) )$ and its perpendicular direction $z$. An isolated N\'eel-type skyrmion has $\Phi =0 $ and thus can be modeled as  \cite{Jiang2017}
 \begin{align} \label{NeelSkyrmionParametrization}
	\vec n 
	&=  \sin \Theta (\rho)  \hat \rho + \cos \Theta (\rho) \hat z \;, 
\end{align}
with a parametrization 
 \begin{align} \label{SmoothSkyrmionModel}
	\Theta(\rho) = \left\{ \begin{matrix}
		\pi \;, & \quad \;  & \rho -P < - \omega  \; \\ \vspace{-0.1in}
		& & \\
		\frac{1}{2} \pi - \frac{\rho - P }{2\omega} \pi \;,   & & - \omega   \leq \rho -P \leq  \omega  \;, \\  \vspace{-0.1in}
		& & \\
		0 \;, &  & \rho -P >  \omega  \;
	\end{matrix}
	\right.
\end{align}
where $P$ and $2\omega$ represent the position and width of the domain wall (DW) illustrated in Fig. \ref{fig:SkyrmionModel}. This DW describes the interpolating region of the negative-charge skyrmion with spin up $\vec n = \hat z$ in the outer region $\rho  > P+ \omega $ and with spin down $\vec n =- \hat z$ in the inner region $\rho  < P- \omega $. 

The continuous skyrmion model enables us to compute the Magnus and dissipative force tensors defined in \eqref{GFactor}. The former is $ \mathcal G_{ij} = \epsilon_{ij} Q =\pm 4\pi  \epsilon_{ij} $. For $\mathcal D_{ij}$, we first compute $\partial_x \vec n = (\partial_x \rho) \Theta'(\rho)  \big(  \cos \Theta \hat \rho - \sin \Theta \hat z \big) -  (\partial_x \phi) \sin \Theta \hat \phi $, where $'$ is $\rho$ derivative. Thus $(\partial_x \vec n)^2 = \cos^2 \phi (\Theta'(\rho))^2 + (\sin^2 \phi / \rho^2) \sin^2 \Theta $. These two terms only contribute for the DW region upon integration and depend on the ratio $\omega /P $. We omit the second term which is much smaller than the first, especially for the skyrmions with a skinny DW region, $\omega /P \ll 1$. Thus, 
	\begin{align}
	\begin{split}
		\mathcal D_{xx}  &=- \int_{P- \omega} ^ {P+ \omega } \hspace{-0.15in} d\rho \rho \int_0^{2\pi}   \hspace{-0.1in} d\phi ~ \frac{\pi^2}{4\omega^2} \cos^2 \phi =- \frac{\pi^3}{4} \frac{ P}{\omega}    \;.
	\end{split}
\end{align}
The diagonal components are the same $\mathcal D_{yy} = \mathcal D_{xx} \equiv D$ even with the omitted term. It is straightforward to check as $(\partial_y \vec n)^2 = \sin^2 \phi (\Theta'(\rho))^2 + (\cos^2 \phi / \rho^2) \sin^2 \Theta $, which gives the same result upon $\phi$ integral. 

Now we check that the off diagonal components vanish, $\mathcal D_{xy} = \mathcal D_{yx}=0$. It is easy to see as $(\partial_x \vec n) \cdot (\partial_y \vec n) = ((\Theta'(\rho))^2 - \sin^2 \Theta/\rho^2 ) \sin \phi \cos \phi $, which vanishes upon $\phi$ integral. Thus, we demonstrate the decomposition $	\mathcal D_{ij} = D_s \delta_{ij} + D_a \epsilon_{ij}$ along with $D_a =0$. 
 \begin{align} \label{ExplicitComputationD}
	\mathcal D_{ij} = D \,  \delta_{ij} = - \frac{\pi^3}{4} \frac{P}{\omega } \;. 
\end{align}
Note \eqref{HallAngleOriginal} gives $ \tan \theta_H = Q/D$ for $R=0$ that reproduces the known fact that the positive- and negative-charge skyrmion Hall angles are the same with opposite signs.   

One can also consider the Bloch-type skyrmions that has the parametrization $ \Phi (\phi) = m \phi + \phi_0$, where $m$ is an integer representing the winding number and $\phi_0$ a constant phase. Similar computation, for example, gives $(\partial_x \vec n)^2 = \cos^2 \phi (\Theta'(\rho))^2 + (m+1)^2 (\sin^2 \phi / \rho^2) \sin^2 \Theta $. Explicitly, it can be shown to $\mathcal D_{xx} =\mathcal D_{yy} = D$ and $\mathcal D_{xy} =\mathcal D_{yx} = 0$. Thus, the dissipative tensor have the same results as in \eqref{ExplicitComputationD}. 

For the rest of the paper, we focus on investigating the consequences of the transverse velocity effect $R$ in the Hall angle \eqref{HallAngleGeneral}. 
 \begin{align} \label{HallAngleSkyrmion}
	\begin{split}
		&\tan \theta_H = \frac{ Q + \alpha D R}{\alpha D - QR } \;.
	\end{split}
\end{align}
Before considering some physical systems, we note that $\tan \theta_H = R $ for $Q=0$. This tells that the positive- and negative-charge skyrmions have the same Hall angles (same transverse motion), for example, in anti-ferromagnetic materials. \\

\noindent {\bf First experimental data sets}  \\  
There exist only a small number of systematic experimental data sets on Hall angles performed for both positive- and negative-charge skyrmions with the same environment \cite{Jiang2017}\cite{VanishingSkyrmionHall}. First, we note that these available data sets consistently show significant, $6.5 - 12\%$, differences. 

Let us start with \cite{VanishingSkyrmionHall} as its data set is clearer to analyze for our purpose. The Hall angles were measured for N\'eel-type half skyrmions ($Q=\pm 2\pi$) in the ferrimagnetic GdFeCo/Pt films by utilizing the spin orbit torque (SOT) technique. The data set at temperature $T=343 \, K$ is given as 
	\begin{align} \label{DataAt343K}
	\theta_{Q>0} = - 35^o \;,  \qquad  \theta_{Q<0} = 31^o \;,
\end{align}
with their $\% \, \text{difference}$ as   
\begin{align}
	\% \, \text{difference} = 12 \% \;.
\end{align}
This is a strikingly large difference with naive expectation that the Hall angles between positive- and negative-charge skyrmions should be the same. 

In \cite{Jiang2017}, the Hall angles for N\'eel-type positive- and negative-charge skyrmions were measured for Ta/CoFeB/TaO$_x$ material, which shows strong pinning potential due to randomly distributed defects. When a ferromagnetic layer is placed on top of a heavy metal layer, polarized electric currents along the heavy metal layer can be used to pump spins into the ferromagnetic layer through the spin Hall effect. While the original data sets, figure $3\bf{c}$ in \cite{Jiang2017}, were collected for different magnetic fields $B$, careful interpolation was performed to obtain data at $B=\pm 5.0 \, Oe $ that are presented in a table \cite{Kim:2020piv}. The saturated Hall angles are $|\theta_{Q>0}| = 31.6^o $ and $\theta_{Q<0} = 29.3^o $ with their $\% \, \text{difference} = 7.6 \%$ for a pulse current. For the opposite current, $\% \, \text{difference} = 5.5 \%$. The combined, conservative, estimate for the difference between the positive- and negative-charge skyrmion Hall angles is $ 6.5\%$. 

It is surprising to observe that there exist a large difference between the positive- and negative-charge skyrmion Hall angles (measured in the same environmental setup). Moreover, the former is consistently bigger than the latter. The original Hall angle formula \eqref{HallAngleOriginal} with $D_a=0$ is not suitable to describe these angles together. Here we would like to use \eqref{HallAngleSkyrmion} to estimate two parameters $ \alpha D$ and $R$. These in turn can reveal the relative strength between the two transverse forces proportional to $ Q $ and $\alpha D R $, which are two numerical terms in \eqref{HallAngleSkyrmion}. 

We start with $\theta_{Q>0} = - 35^o$ for $Q=2\pi$ and $\theta_{Q<0} = 31^o $ for $Q=-2\pi$ \cite{VanishingSkyrmionHall}. By using \eqref{HallAngleSkyrmion} twice, we can estimate $ \alpha D = -9.68$ and $R=-0.035$, where we use $\alpha D <0$. Thus the new contribution with $R$ is estimated to be 
	\begin{align} \label{NewHallContributionA}
	\frac{\alpha D R}{Q} = 5.4 \% \;.
\end{align}
Thus the transverse force due to the new transverse effect $R$ is about $5.4 \%$ of the conventional skyrmion Hall effect, which is surprisingly large. 

With the estimated values $\alpha D $ and $R$, we revisit the skyrmion Hall angle  \eqref{HallAngleSkyrmion} to see the effects of the new terms introduced in \eqref{ThieleEQ}. For a positive-charge skyrmion $Q>0$, the new term $ \alpha D R >0 $ enhances the transverse motion of the skyrmion, while the other new term $QR<0$ reduces its longitudinal motion. For a negative-charge skyrmion, the effect is opposite. 

We turn to the data \cite{Jiang2017}, analyzed in detail \cite{Kim:2020piv}, $\theta_{Q>0} = - 31.6^o$ for $Q=4\pi$ and $\theta_{Q<0} = 29.3^o $ for $Q=-4\pi$ for a current with $B= \pm 5 \, Oe$. It is estimated to $ \alpha D = -21.4$ and $R=-0.020$, which give $\alpha D R / Q = 3.4 \%$. Another data set, $\theta_{Q>0} = - 32^o$ for $Q=4\pi$ and $\theta_{Q<0} = 30.3^o $ for $Q=-4\pi$, for the opposite current gives $\alpha D R / Q = 2.5 \%$. The combined estimate is $\alpha D R / Q = 2.9 \%$.

Thus we conclude that, from two systematic experimental data sets, there exist an unexpected large difference: the positive-charge skyrmion Hall angle is $6.5 - 12 \%$ larger than the negative-charge skyrmion Hall angle. The corresponding transverse force amounts to $2.9 - 5.4\%$ of the conventional Hall effect due to the skyrmion charge. These were analyzed before in \cite{Kim:2020piv}.   \\

\begin{figure}[t!]
	\begin{center}
		\includegraphics[width=0.43\textwidth]{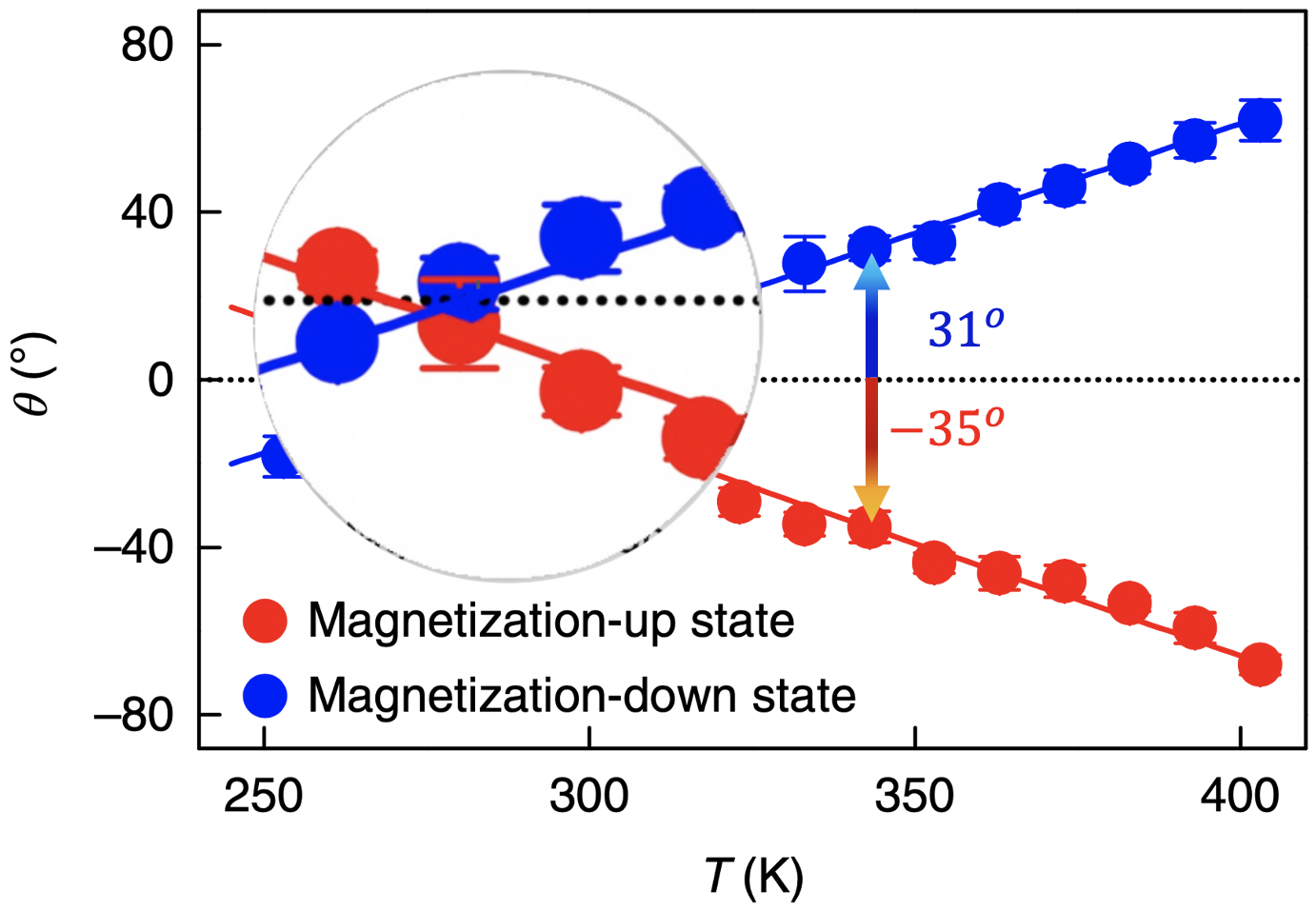} 
		\caption{\footnotesize\small  Systematic Hall angle data \cite{VanishingSkyrmionHall} with enlargement around the angular momentum compensation temperature and the data set at $T=343 \, K$. Magnetization-up state is identified as half skyrmion in \cite{VanishingSkyrmionHall} and has $Q=-2\pi$ in our convention.   \vspace{-0.2in} 
		}
		\label{fig:HallAnglesCompensationPointA}
	\end{center}
\end{figure}

\noindent {\bf Second experimental data set}  \\  
We return to \cite{VanishingSkyrmionHall} as it has more interesting data, FIG. \ref{fig:HallAnglesCompensationPointA}, for the positive- and negative-charge skyrmion Hall angles across the angular momentum compensation temperature $T_{A} \approx 283 ~K$ for a rare earth–3d-transition metal (ferrimagnetic) compound, GdFeCo/Pt film. The angular momentum compensation point is defined as the temperature when the spin densities of the two sub-compounds cancel each other as discussed below. For a clearer presentation, we also define another temperatures $\tilde T_\pm$, where the positive- and negative-charge skyrmion Hall angles vanish.

Before considering the Hall angle formula, we note interesting observations for FIG. \ref{fig:HallAnglesCompensationPointA},: (i) the positive-charge skyrmion Hall angles for $T> T_{A} $ are consistently larger than the negative-charge skyrmion angle as advertised above, (ii) the positive-charge skyrmion compensation temperature $(\tilde T_{+})$ is smaller than the negative-charge  skyrmion one $(\tilde T_{-})$, and (iii) when the two Hall angles coincide, happening at $\tilde T_0$ ($\tilde T_{+} < \tilde T_0 < \tilde T_{-}$), their values are negative, not vanishing. Thus, we can see that the Hall angle lines form a triangle with three vertices located at the three temperatures $\tilde T_{+}, \tilde T_0$ and $\tilde T_{-}$. While the data are within the experimental uncertainties, the significant difference between the positive- and negative-charge skyrmion Hall angles deserves to investigate these observations further. More precision experiments will surely clarify these issues. The rest of the paper is devoted to look into them from a theoretical stand point. 

Ferrimagnet compound is composed with two subnetworks of magnetization, $\vec M_1 $ and $\vec M_2$, that are anti-ferromagnetically coupled through an effective local exchange field. While the gyromagnetic ratios $\gamma_a = M_a/s_a ~(a=1,2)$ are constants as temperature changes, their magnet moments $\vec M_a$ and spin densities $\vec s_a$ of the two subnetworks change differently. There are two special temperatures, the magnetization compensation point where the net magnetization vanishes and the angular momentum compensation point ($T= T_A$), across which the net spin density $ s_n (T_A) = s_1 (T_A) - s_2 (T_A) =0$ changes its sign. Its general treatments using two independent magnetization vectors were developed in  \cite{FerrimagnetFastMovongDomainWall}\cite{FerrimagnetTheory1}. We note that the net spin density increases as temperature decreases.

When the exchange field is sufficiently large, the two magnetization vectors remain strongly coupled and anti-parallel to each other \cite{MMFerrimagnetBook}. Thus, $\vec \mu = \vec \mu_1 = - \vec \mu_2$ using the unit vectors $\mu_a$ as $\vec M_a = M_a \vec \mu_a$. We define the net magnetic moment $M_n = M_1 - M_2 $. The dynamics of $\vec \mu$ of the combined system can be described as 
\begin{align} \label{FerrimagnetOEq3}
	\begin{split}
		s_n \dot{\vec \mu} = - M_n \vec \mu \times \vec H_{\text{eff}} + s_t \alpha_{\text{eff}} \vec \mu \times \dot{\vec \mu} \;,
	\end{split}
\end{align}
where the net and total spin densities are $ s_n = s_1 - s_2$ and $s_t = s_1+s_2 $, respectively. $\alpha_{\text{eff}}= (\alpha_1 s_1 + \alpha_2 s_2 )/( s_1 +  s_2)$ is the effective damping parameter. 

We are ready to apply our generalized Hall angle \eqref{HallAngleSkyrmion} for the experimental data in FIG. \ref{fig:HallAnglesCompensationPointA}. \eqref{FerrimagnetOEq3} is simpler, yet consistent with the model adapted in \cite{VanishingSkyrmionHall}, which used the Magnus force as $ \vec F_g = s_n Q (\hat z \times \vec v)$ and the viscose force as $ \vec F_d = s_t \alpha D \vec v$. Thus, the spin (and temperature) dependence can be incorporated by $Q \to s_n Q$ and $\alpha D \to s_t \alpha D $. Thus, 
 \begin{align} \label{SkHangleSnet}
 	\tan \theta_{H}= \frac{s_n Q+ s_t \alpha D R}{s_t \alpha D - s_n QR } \;. 
 \end{align}
When $R=0$, $\tan \theta_{H}= s_n Q / s_t \alpha D $, which vanishes at the compensation temperature as $s_n (T_A) =0$. The positive- and negative-charge skyrmion Hall angles are the same (with opposite signs).  

We highlight the Hall angle \eqref{SkHangleSnet}, among its interesting details, with its dependence on the spin densities. It only depends on the combination $s_n/s_t$, which turns out to be a monotonically decreasing function of temperature $T$. By analyzing the data \eqref{DataAt343K} at $T=343 \, K$, we arrive the same estimates $(s_t/ s_n) (\alpha D R/ Q) = 5.4\%$ in \eqref{NewHallContributionA}, along with $R = -0.035$ and $(s_t/s_n)(\alpha D) = -9.68$ \cite{footnote}.

Let us look at the Hall angle \eqref{SkHangleSnet} more carefully near the compensation point as a function of $ s_n/s_t$. 
\begin{itemize}
		\item[a.] When $ s_n = 0$, that happens $T= T_A $,  
	\begin{align}  
		\tan \theta_{H}= R \;. 
	\end{align}
	The two Hall angles are the same at $T = \tilde T_0$ and negative as $R<0$. Thus $\tilde T_0 $ coincide with the compensation temperature $T_A$. 
	
	\item[b.] The positive- and negative-charge skyrmion Hall angles vanish at different temperatures as they happen when $ s_n Q+ s_t \alpha D R = 0$. In terms of spin densities, 
	\begin{align}  
		\frac{s_n}{s_t} \Big|_{T= \tilde T_{\pm}} = - \frac{\alpha D R}{Q}  \;. 
	\end{align}
	Note also {\it vanishing Hall angle temperatures} $\tilde T_{\pm}$ do not coincide with the compensation temperature $T_{A}$. The positive-charge skyrmion Hall angle vanishes when $s_n/s_t >0 $ (thus $\tilde T_{+} < T_A$), while the negative-charge skyrmion one vanishes when $s_n/s_t <0 $ (thus $\tilde T_{-} > T_A$), where we use $\alpha D>0 $ and $R<0$ \cite{footnote}. 

\end{itemize}
From a) and b), we confirm that these three different temperatures ($\tilde T_{A+} < \tilde T_0 < \tilde T_{A-}$) serve as vertices of the triangle mentioned above for FIG. \ref{fig:HallAnglesCompensationPointA}.

\begin{figure}[t!]
	\begin{center}
		\includegraphics[width=0.45\textwidth]{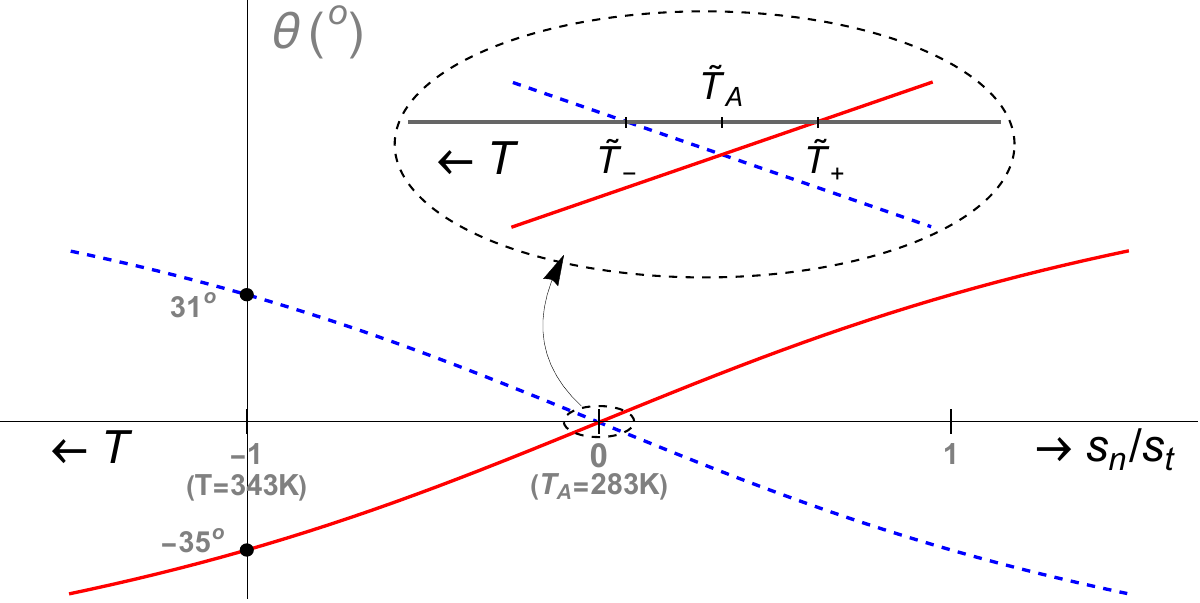} 
		\caption{\footnotesize\small Plot generated for $\theta_{H}$ given in \eqref{SkHangleSnet} as a function of $s_n/s_t $, which is a decreasing function of $T$. Solid red line is for skyrmion, while dashed blue for antiskyrmion. This plot captures the essential features of the experimental data presented in figure 4 in \cite{VanishingSkyrmionHall}. 
			\vspace{-0.2in} 
		}
		\label{fig:TrangleEnlarged}
	\end{center}
\end{figure}

Finally, we plot the positive-charge (red) and negative-charge (blue) skyrmion Hall angles in FIG. \ref{fig:TrangleEnlarged} as a function of $s_n/s_t$ for the range $ -1.5 \leq s_n/s_t \leq 1.5$. Here we use constant values for $R$ and $ \alpha D$ obtained at the reference point $s_n/s_t =-1 $ at $T=343 \, K$. We note that this plot captures essential features of the experimental Hall angle data as a function of spin density that is presented in figure 4 of \cite{VanishingSkyrmionHall}. In addition, the inset of FIG. \ref{fig:TrangleEnlarged}, that manifests the triangle near the compensation point, is consistent with FIG. \ref{fig:HallAnglesCompensationPointA}. \\

\noindent {\bf What accounts for the transverse velocity?} \\
In this paper we have generalized the Thiele equation with the transverse velocity effect introduced in \eqref{SkyrmionCM} and demonstrated its usefulness for describing the surprising mismatches ($6.5 - 12 \%$) between the positive- and negative-charge skyrmion Hall angles. In particular, this new transverse force accounts for $2.9 - 5.4 \%$ of the conventional skyrmion Hall effect. Moreover, our generalized Hall angle formula \eqref{SkHangleSnet} (also \eqref{HallAngleSkyrmion}) reproduces a non-trivial experimental Hall angle data in the context of the ferrimagnetic compound near the compensation point. See FIG. \ref{fig:TrangleEnlarged}. 
   
One can ask whether there is a physical motivation for introducing the transverse velocity component $R$ to the Thiele equation. The answer is affirmative. 

Recent study of Hydrodynamics revealed that there exists a universal transport coefficient, Hall viscosity \cite{Avron:1995}, in the absence of mirror (parity) symmetry \cite{Jensen:2011xb}\cite{Bhattacharya:2011tra}. It has been extensively studied theoretically in quantum Hall systems and linked to a half of the system's angular momentum density \cite{Read:2008rn} or Hall conductivity \cite{Hoyos:2011ez}. More systematic relations were studied in \cite{Bradlyn:2012ea}\cite{Hoyos:2014lla}. 

This Hall viscosity was introduced to skyrmion motion, whose systems also have broken parity, by utilizing the topological Ward identity \cite{Kim:2015qsa}\cite{Kim:2019vxt}\cite{KimBook}. It is transverse to the skyrmion's motion and dissipationless. Moreover, the analysis of Kubo formula tells that Hall viscosity does not depend on skyrmioin charges and thus pushes the positive- and negative-charge skyrmions toward the same transverse direction \cite{Kim:2020piv}. Once added to the skyrmion Hall effect, the positive-charge skyrmion Hall angle is bigger than that of the negative-charge skyrmion.Thus, we can conclude the effect of Hall viscosity is estimated to be $2.9 - 5.4 \%$ of the conventional skyrmion Hall effect according to our analysis. Precision skyrmion experiments focused on the positive- and negative-charge skyrmion Hall angles are an ideal ground for reliably confirming the existence of Hall viscosity!  \\

\noindent {\it Acknowledgments:} 
This work has been partially supported by Wisys Spark grant and UW-Parkside summer research funds. We thank to the anonymous referees' comments, which improve the paper's clarity.


\begin{thebibliography}{999}
\bibitem{footnoteSkyrmionCharge} 
	Here we use a slightly different definition for skyrmion charge with an extra $4\pi$ compared to \cite{SKCharge}. Moreover, we point out that there are different conventions for defining  skyrmion and antiskyrmion as illustrated in Fig. 2.4 of \cite{Han} and explained there. To remove further confusions, we use the term positive- and negative-charge skyrmions when it is necessary to distinguish them. For comparison below, we also mention that the Magnetization-up state defined in Fig 3 and Fig.4 of \cite{VanishingSkyrmionHall} has $Q=+1/2$.
\bibitem{footnote} 
	We note $s_n/s_t=-1$ at $T= 343 \, K$ in our convention. To analyze Hall angles and other quantities as a function of $s_n/s_t$, we need to set $ R = -0.035$ and $ \alpha D = 9.68$. This is consistent with the estimates near \eqref{NewHallContributionA}.
\end{thebibliography}
\end{document}